\UseRawInputEncoding
\documentclass[prx, notitlepage, twocolumn, nofootinbib, superscriptaddress]{revtex4-1}
\usepackage[colorlinks=true,citecolor=blue,linkcolor=blue]{hyperref}
\usepackage{graphicx}
\usepackage{color}
\usepackage{dcolumn}
\usepackage{bm}
\usepackage{mathrsfs}
\usepackage{amsmath}
\usepackage{amssymb}
\usepackage{amsthm}
\usepackage{amsfonts}
\usepackage{enumerate}
\usepackage{latexsym}
\usepackage{hyperref}
\usepackage{centernot}
\usepackage{multirow}
\usepackage[caption=false]{subfig}

\begin{document}

\title{First-principles study of spin orbit coupling contribution to
anisotropic magnetic interaction}

\begin{abstract}
Anisotropic magnetic exchange interactions lead to a surprisingly rich
variety of the magnetic properties. Considering the spin orbit coupling
(SOC) as perturbation, we extract the general expression of a bilinear spin
Hamiltonian, including isotropic exchange interaction, antisymmetric
Dzyaloshinskii-Moriya (DM) interaction and symmetric $\Gamma $ term. Though
it is commonly believed that the magnitude of the DM and $\Gamma $
interaction correspond to the first and second order of SOC strength $%
\lambda $ respectively, we clarify that the term proportional to $\lambda
^{2}$ also has contribution to DM interaction. Based on combining magnetic
force theorem and linear-response approach, we have presented the method of
calculating anisotropic magnetic interactions, which now has been
implemented in the open source software WienJ. Furthermore, we introduce
another method which could calculate the first and second order SOC
contribution to the DM interaction separately, and overcome some
shortcomings of previous methods. Our methods are successfully applied to
several typical weak ferromagnets for $3d$, $4d$ and $5d$ transition metal
oxides. We also predict the conditions where the DM interactions
proportional to $\lambda $ are symmetrically forbidden while the DM interactions
proportional to $\lambda ^{2}$ are nonzero, and believe that it is
widespread in certain magnetic materials.
\end{abstract}

\date{\today }
\author{Di Wang}
\affiliation{National Laboratory of Solid State Microstructures and School of Physics,
Nanjing University, Nanjing 210093, China}
\affiliation{Collaborative Innovation Center of Advanced Microstructures, Nanjing
University, Nanjing 210093, China}
\author{Xiangyan Bo}
\affiliation{Nanjing University of Posts and Telecommunications, Nanjing 210023, China}
\author{Feng Tang}
\affiliation{National Laboratory of Solid State Microstructures and School of Physics,
Nanjing University, Nanjing 210093, China}
\affiliation{Collaborative Innovation Center of Advanced Microstructures, Nanjing
University, Nanjing 210093, China}
\author{Xiangang Wan}
\thanks{The corresponding author: xgwan@nju.edu.cn.}
\affiliation{National Laboratory of Solid State Microstructures and School of Physics,
Nanjing University, Nanjing 210093, China}
\affiliation{Collaborative Innovation Center of Advanced Microstructures, Nanjing
University, Nanjing 210093, China}
\maketitle


\section{Introduction}

Magnetic properties can be typically described by a quadratic spin
Hamiltonian, which is the basis of most magnetic theoretical investigations
\cite{dm1book,book-2,book-exchange}. Generally, spin-orbit coupling (SOC)
always exists and leads to the anisotropic magnetic interactions with low
symmetry. The general form of the bilinear expression of a spin exchange
Hamiltonian could be written as

\begin{eqnarray}
H &=&\sum_{i<j}J_{ij}\mathbf{S}_{i}\cdot \mathbf{S}_{j}+\sum_{i<j}\mathbf{D}%
_{ij}\cdot \lbrack \mathbf{S}_{i}\times \mathbf{S}_{j}]+\sum_{i<j}\mathbf{S}%
_{i}\cdot \Gamma _{ij}\cdot \mathbf{S}_{j}  \notag \\
&&  \label{eq-1}
\end{eqnarray}

where the first term describes the isotropic Heisenberg Hamiltonian, the
second one represents the Dzyaloshinskii-Moriya (DM) \cite{DM-D,DM-M,dmi2021}
interaction, and the third one is marked as $\Gamma $ term \cite{DM-M}. The
antisymmetric DM interaction, which comes from the combination of low
symmetry and SOC, is introduced by Dzyaloshinskii \cite{DM-D} and Moriya
\cite{DM-M} in a phenomenological model and a microscopic model
respectively. It is commonly believed that DM
and $\Gamma $ term are contributed by first and second order of SOC
respectively \cite{dm1book,DM-M}. Generally, DM interaction favors twisted
spin structures and is constrained by the crystal symmetry. For example,
when a inversion center located at the bond center\ of two magnetic ion
sites, the DM interaction between these two magnetic ions should be zero due
to its antisymmetric property \cite{DM-M,dmi2021}. Now the DM interaction is
invoked to explain numerous interesting magnetic systems featuring
non-collinear spin textures, such as weak ferromagnets \cite{DM-D,DM-M},
helimagnets \cite{helical}, skyrmion formation \cite%
{skyrmion-0,skyrmion,skyrmion-1} and chiral domain walls \cite%
{domain-0,domain}. In addition, the DM interaction also plays an important
role in multiferroic materials \cite%
{mferro-0,mferro-1,mferro-2,mferro,wang2009multiferroicity}, topological
magnon materials \cite{magnon-dm-1,magnon-dm-2,magnon-dm-3} and spintronics
\cite{wang2018topological}. It is worth mentioning that, since DM
interactions are very sensitive to small atomic displacements and symmetry
restrictions, it can also be used to reveal the interplay of delicate
structural distortions and complex magnetic configurations \cite{FeGe}.

Recently, the first-principles study of magnetic exchange interactions
especially DM interaction, has also attracted much interest \cite%
{book-exchange,mappingreview,xianghongjunDM,mywork-1,mywork-2,J-spiral,dm-spiral,dm-add1,dm-add2,dm-spiral-2,dm-szadd-5,J-1987,bruno-2003,dm-2000,dm-2005,dm-2010,dm-add3,dm-add4,dm-add5,dm-add9,dm-add10,dm-szadd-2,dm-szadd-3,dm-szadd-6,dm-szadd-1,dm-x,dm-current1,dm-current2,Berry}%
. A popular numerical method is the energy-mapping analysis \cite%
{book-exchange,mappingreview,xianghongjunDM} to estimate magnetic
interactions from the energy differences of various magnetic structures.
However, this approach becomes inconvenient for the complicated systems
where it is not clear how many exchange interactions needs to be considered,
since in some magnetic compounds the magnetic moments may couple over a
variety of distances, and even the ninth-nearest-neighbor coupling plays an
important role \cite{mywork-1,mywork-2}. Meanwhile, in itinerant magnetic
systems, the magnetism is not so localized and\ the calculated magnetic
moments\ may depend on the magnetic configurations, which also significantly
affects the accuracy of the calculated DM interactions. Another approach
using total energy differences could extract DM strength by directly
calculating the energies of spirals with the finite vector $q$ \cite%
{J-spiral,dm-spiral,dm-add1,dm-add2,dm-spiral-2,dm-szadd-5}. Meanwhile, an
efficient approach is proposed based on the magnetic force theorem \cite%
{J-1987,bruno-2003,dm-2000,dm-2005,dm-2010,dm-add3,dm-add4,dm-add5,dm-add9,dm-add10,dm-szadd-2,dm-szadd-3,dm-szadd-6,dm-szadd-1}%
. Katsnelson $et$ $al.$ \cite{dm-2000} have derived the expression for DM
interaction term based on Green's function approach. This method is applied
to a large number of magnetic materials such as the antiferromagnets with
weak ferromagnetism \cite{dm-2005}, thin magnetic films \cite{dm-add3},
diluted magnetic semiconductors \cite{dm-add4} and various other magnetic
materials \cite{dm-add9,dm-add10,dm-szadd-2,dm-szadd-3,dm-szadd-6}. This
Green's function approach was previously formulated in first-principles
codes with direct definition of a localized orbitals basis set such as
linear muffin-tin orbitals method \cite{lmto}. Furthermore, they have also
developed the method of calculating DM interactions using Wannier function
formalism \cite{dm-2010,dm-add5}. In addition, DM interactions could also be
estimated by computing the long-wave length limit of the spin susceptibility
\cite{dm-x}, the expectation value of the spin current density \cite%
{dm-current1,dm-current2}, or utilizing Berry phase \cite{Berry}.

In this paper, up to the second-order perturbation of SOC, we revisit the
general expression of anisotropic magnetic interactions. We clarify that the
second order term of SOC has contribution to both the antisymmetric DM
interaction and symmetric $\Gamma $ interaction, and reveal that their
distinction arises from different hopping processes as shown in Fig. \ref%
{toymodeldm} and following. Note that in these approaches using total energy
differences, such as energy-mapping \cite%
{book-exchange,mappingreview,xianghongjunDM}, spirals approach \cite%
{dm-spiral,dm-add1} and the approaches through calculating energy variations
due to spin rotations \cite{dm-2000,wan2006,wan2009}, can get the entire DM
interaction but do not distinguish the contribution from first or second
order of SOC. On the other hand, one can only consider DM interactions with
the first order of SOC in their perturbation\ schemes \cite%
{dm-2005,dm-current2}. We extend the method of calculating Heisenberg
interactions based on combining magnetic force theorem and linear-response
approach \cite{J-1987,wan2006,wan2009,wan2011,wan2021,mywork-1} to estimate
DM and $\Gamma $ interactions, and the algorithm of our proposed method is
now implemented in the open source called WienJ \cite{wienj}, as an
interface to the linearized augmented plane wave (LAPW) software Wien2k \cite%
{wien2k}. Furthermore, to overcome some shortcomings of previous methods, we
develop a new method that can estimate the first and second order SOC
contribution to the DM exchange couplings separately. While our methods can
also calculate $\Gamma $ interaction, here we only present the results of
Heisenberg and DM interactions since they could be compared with many
previous works. We have applied our methods to several representatives of\
canted antiferromagnetic materials La$_{2}$CuO$_{4}$, Ca$_{2}$RuO$_{4}$ and
Ca$_{3}$LiOsO$_{6}$ for $3d$, $4d$ and $5d$ transition metal oxides, and the
calculation results are consistent with the experiment. Particularly, we
find that the DM interaction proportional to $\lambda ^{2}$ can not be
ignored in $4d$ transition metal oxide Ca$_{2}$RuO$_{4}$, and the DM
interactions proportional to $\lambda $ and $\lambda ^{2}$ have the same
magnitude in $5d$ transition metal oxide Ca$_{3}$LiOsO$_{6}$. As shown in
the following, the DM interactions proportional to $\lambda $ and$\ \lambda
^{2}$ involve different exchange channels. Thus, based on the symmetry
analysis, we explore the possibility that the DM interactions proportional
to $\lambda $ are symmetrically forbidden while the DM interactions
proportional to $\lambda ^{2}$ still exist. We believe that this case is
widespread in certain magnetic materials, and our method would play more
important role in these magnetic systems.

\section{Method}

\subsection{Anisotropic magnetic interactions by perturbation theory}

We start from an effective model:

\begin{eqnarray}
H &=&H_{0}+H_{t}+H_{U}+H_{soc}  \notag \\
&=&\sum_{i\alpha \sigma }\varepsilon _{\alpha }c_{i\alpha \sigma
}^{+}c_{i\alpha \sigma }+\sum_{ij\alpha \beta \sigma }t_{\alpha \beta
}^{ij}c_{i\alpha \sigma }^{+}c_{j\beta \sigma }  \notag \\
&&+U\sum_{i}n_{i,\uparrow }n_{i,\downarrow }+\sum_{i}\lambda l_{i}\cdot s_{i}
\label{hubbard}
\end{eqnarray}

where $H_{0}$, $H_{t}$, $H_{U}$ and$\ H_{soc}$ represent the on-site orbital
energy, the hopping term,\ the Hubbard $U$\ term and the SOC term,
respectively. Here $i,j$ represent the site index, while $\alpha ,\beta $
represent the orbital index and $\sigma $ represents the spin index. We
consider the spin exchange interaction between the magnetic ions located at
site $A$ and site $B$. We label the ground state and the unoccupied states
at site $A$ as $n$ and $m$, respectively. Similarly, the ground state and
the excited states at site $B$ are labeled as $n^{\prime }$ and $m^{\prime }$%
, respectively.

\begin{figure*}[tbph]
\centering\includegraphics[width=0.90\textwidth]{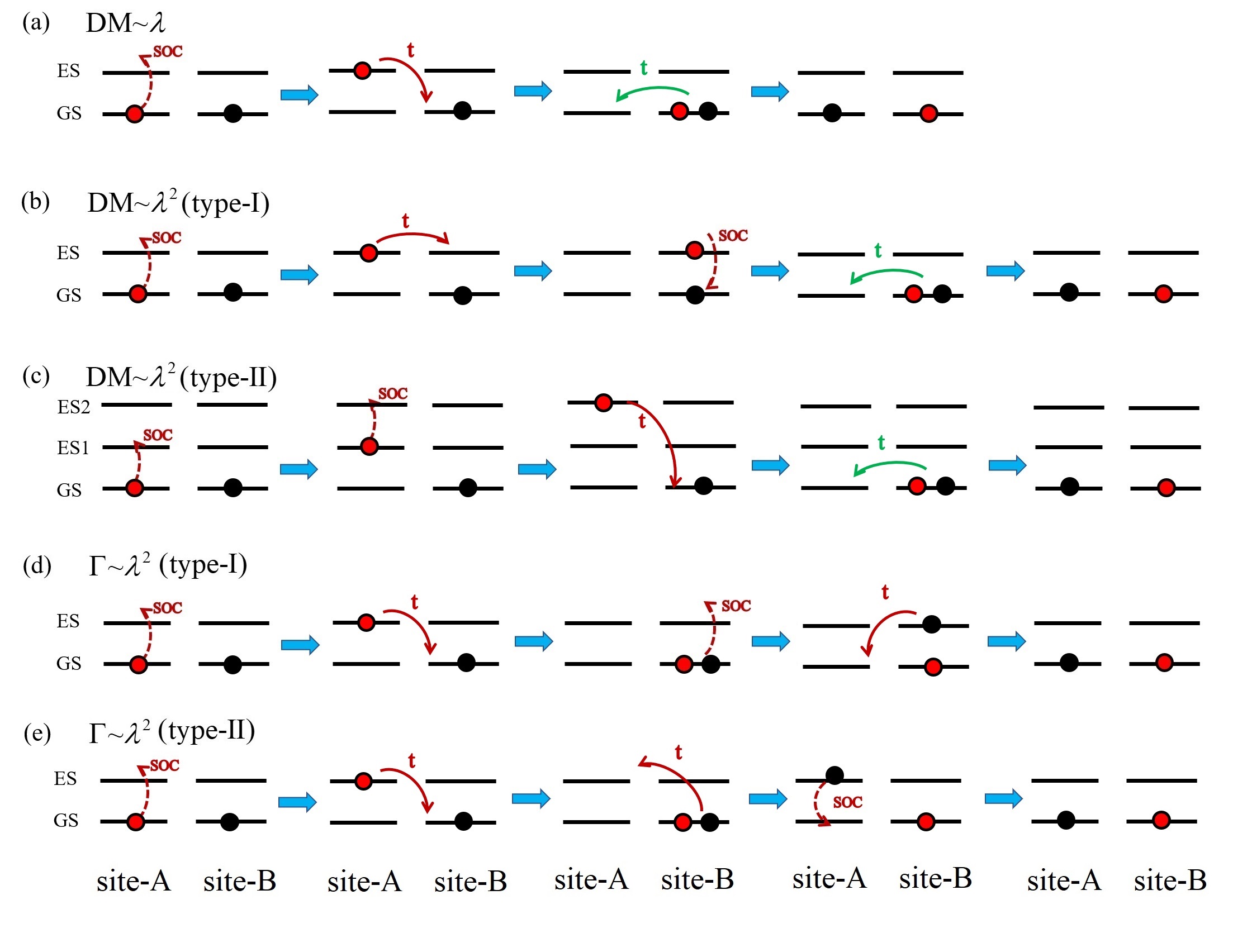}
\caption{Schematic pictures of exchange paths for anisotropic magnetic
interactions between site A and site B. The dotted line represents the SOC
excitation process, while the solid line represents the hopping process. (a)
represents the DM interactions for the first order of SOC. The DM
interactions for the second order of SOC have two types of perturbation
processes (b) and (c). Meanwhile, the perturbation processes for the
symmetric $\Gamma $ term are also shown in (d) and (e) for comparison. Here
GS and ES represent ground state and excited state respectively. It is worth
mentioning that, the perturbation processes of DM interactions involve the
hopping between GSs, which denoted by the green solid line. Meanwhile, $%
\Gamma $ terms would only involve the hopping processes between GS and ES.}
\label{toymodeldm}
\end{figure*}

When SOC is not considered, the Heisenberg interactions $H_{eff}=J\mathbf{S}%
_{A}\cdot \mathbf{S}_{B}$ can be obtained by considering the hopping term as
perturbations for the case of $U>>t$ \cite{anderson1950}. Considering the
SOC term $\lambda l\cdot s$ as perturbation, the first-order SOC
contribution to effective spin model $H_{eff}^{(1)}$ has the expression of
antisymmetric DM interaction as $H_{eff}^{(1)}=\mathbf{D}^{(1)}(\mathbf{S}%
_{A}\times \mathbf{S}_{B})$\ where $\mathbf{D}^{(1)}$ could be written as
\cite{DM-M}
\begin{equation}
\left( D^{\alpha }\right) ^{(1)}=-4i\frac{\lambda t_{nn^{\prime }}}{U}\left(
\sum_{m}\frac{l_{mn}^{\alpha }}{\varepsilon _{m}-\varepsilon _{n}}%
t_{mn^{\prime }}-\sum_{m^{\prime }}\frac{l_{m^{\prime }n^{\prime }}^{\alpha }%
}{\varepsilon _{m^{\prime }}-\varepsilon _{n^{\prime }}}t_{m^{\prime
}n}\right)  \label{dm1jie}
\end{equation}

Meanwhile, by considering perturbation theory up to the second-order SOC
correction (see details in Appendix A), we find that the second-order SOC
correction has the contribution to both DM term $\mathbf{D}^{(2)}$ and the $%
\Gamma $ term, where $\mathbf{D}^{(2)}$ could be written as

\begin{eqnarray}
\left( D^{\alpha }\right) ^{(2)} &=&2\frac{\lambda ^{2}t_{nn^{\prime }}}{U}%
\sum_{m_{,}m^{\prime }}\frac{l_{m^{\prime }n^{\prime }}^{\beta
}l_{mn}^{\gamma }-l_{m^{\prime }n^{\prime }}^{\gamma }l_{mn}^{\beta }}{%
\left( \varepsilon _{m^{\prime }}-\varepsilon _{n^{\prime }}\right) \left(
\varepsilon _{m}-\varepsilon _{n}\right) }t_{mm^{\prime }}  \notag
\label{dm2jie} \\
&&-2\frac{\lambda ^{2}t_{nn^{\prime }}}{U}\sum_{m_{1,}m_{2}}\frac{%
l_{m_{1}m_{2}}^{\beta }l_{m_{2}n}^{\gamma }-l_{m_{1}m_{2}}^{\gamma
}l_{m_{2}n}^{\beta }}{\left( \varepsilon _{m_{1}}-\varepsilon _{n}\right)
\left( \varepsilon _{m_{2}}-\varepsilon _{n}\right) }t_{m_{1}n^{\prime }}
\notag \\
&&+2\frac{\lambda ^{2}t_{nn^{\prime }}}{U}\sum_{m_{1}^{\prime
},m_{2}^{\prime }}\frac{l_{m_{1}^{\prime }m_{2}^{\prime }}^{\beta
}l_{m_{2}^{\prime }n^{\prime }}^{\gamma }-l_{m_{1}^{\prime }m_{2}^{\prime
}}^{\gamma }l_{m_{2}^{\prime }n^{\prime }}^{\beta }}{\left( \varepsilon
_{m_{1}^{\prime }}-\varepsilon _{n^{\prime }}\right) \left( \varepsilon
_{m_{2}^{\prime }}-\varepsilon _{n^{\prime }}\right) }t_{m_{1}^{\prime }n}
\notag \\
&&
\end{eqnarray}

Meanwhile, the expression of parameter $\Gamma ^{(2)}$ could be seen in Eq. (12) of Appendix.

Here we present schematic pictures of the exchange processes in Fig. \ref%
{toymodeldm}. It is easy to see that, the bilinear spin exchange Hamiltonian
should contain two hopping processes between two sites A and B as shown in
Fig. \ref{toymodeldm}. Considering up to the second-order perturbation of
SOC, we find that there are several different exchange processes. Among
them, the first-order SOC correction has only\ DM contribution $\mathbf{D}%
^{(1)}$, as shown in Fig. \ref{toymodeldm}(a), which represents the first
term of Eq. (\ref{dm1jie}). When swap the sites A and B in Fig. \ref%
{toymodeldm}(a), one can obtain the second term of Eq. (\ref{dm1jie}).
Meanwhile, the second-order SOC correction has not only the contribution to\
DM interaction, but also the contribution to\ $\Gamma $ term as shown in
Fig. \ref{toymodeldm}(b)-(e). While the type-I $\mathbf{D}^{(2)}$ as shown
in Fig. \ref{toymodeldm} (b) represents the first term of Eq. (\ref{dm2jie}%
), the type-II $\mathbf{D}^{(2)}$ as shown in Fig. \ref{toymodeldm}(c)
represents the second term of Eq. (\ref{dm2jie}), and the third term of Eq. (%
\ref{dm2jie}) could be obtained by swapping the sites A and B. It is worth
mentioning that, the exchange processes for DM interactions (Fig. \ref%
{toymodeldm}\ (a)-(c)) involve the hopping between ground states, which
denoted by the green solid line in Fig. \ref{toymodeldm} and $t_{nn^{\prime
}}$ in Eq. (\ref{dm1jie})(\ref{dm2jie}) respectively. In sharp contrast,
there are only\ hoppings between ground states and excited states in the
processes of $\Gamma $ terms as shown in Fig. \ref{toymodeldm} (d)-(e).
Therefore, we emphasize that the essential difference between DM and $\Gamma
$ term is from their different hopping processes, rather than the commonly
believed different orders of SOC \cite{dm1book,DM-M}.

\subsection{Magnetic interactions in the first-principles approach}

Firstly we present the method to calculate magnetic interactions based on
the force theorem and linear-response approach \cite{J-1987,wan2006,wan2009}%
, which could be written as the following form \cite{wan2006}

\begin{eqnarray}
J_{\mathbf{R}_{l}+\boldsymbol{\tau },\mathbf{R}_{l^{\prime }}+\boldsymbol{%
\tau }^{\prime }}^{\alpha \beta } &=&\sum_{nkn^{\prime }k^{\prime }}\frac{%
f_{nk}-f_{n^{\prime }k^{\prime }}}{\varepsilon _{nk}-\varepsilon _{n^{\prime
}k^{\prime }}}\left\langle \psi _{nk}\left\vert \left[ \sigma \times B_{%
\boldsymbol{\tau }}^{{}}\right] _{\alpha }\right\vert \psi _{n^{\prime
}k^{\prime }}\right\rangle  \notag \\
&&\times \left\langle \psi _{n^{\prime }k^{\prime }}\left\vert \left[ \sigma
\times B_{\boldsymbol{\tau }^{\prime }}^{{}}\right] _{\beta }\right\vert
\psi _{nk}\right\rangle e^{i(k^{\prime }-k)(\mathbf{R}_{l}-\mathbf{R}%
_{l^{\prime }})}  \notag \\
&&  \label{calj}
\end{eqnarray}

This method has been successfully applied to calculate Heisenberg interactions in
various magnetic materials \cite%
{J-1987,wan2006,wan2009,wan2011,wan2021,mywork-1}. Considering the case of $%
\alpha \neq \beta $, we extend this method to estimate DM and $\Gamma $
interactions, and the algorithm of this method is now implemented in the
open source called WienJ \cite{wienj}, as an interface to Wien2k \cite%
{wien2k}. It is worth mentioning that, the general expression of bilinear
spin exchange parameter $J_{{}}^{\alpha \beta }$, which could be written in $%
J$, $D$ and $\Gamma $ as Eq. (\ref{eq-1}), has 9 independent components.
However, one can only yield\ 4 out of 9 components of $J_{{}}^{\alpha \beta
} $ for a given magnetic configuration. For example, for the collinear
magnetic configuration with all spin moments lying along the $z$-axis, only
the four spin exchange parameters $J_{{}}^{xx}$, $J_{{}}^{yy}$\ , $%
J_{{}}^{xy}$\ and $J_{{}}^{yx}$ can be estimated. Therefore, to obtain the
full nine spin exchange parameters $J_{{}}^{\alpha \beta }$ (i.e. $J$, $D$
and $\Gamma $ terms), one need perform different first-principles self-consistent
calculations for at least three independent orientations of the
magnetization \cite{mywork-1}. Based on the self-consistent results from
different spin orientations, the magnetic interactions could be calculated
from Eq. (\ref{calj}) \cite{mywork-1}. However, these self-consistent
calculations by choosing three different spin orientations would produce 12
parameters, resulting in that sets of parameters $J^{\alpha \beta }$ are not
necessarily unique, and naturally leading to the calculation\ deviation.

To reduce this calculation\ deviation, we also propose a new method when SOC
is relatively small. Firstly, we perform the standard LSDA (+$U$)
calculations. Based on the eigenvalues $\varepsilon _{nk}$ and eigenstates $%
\psi _{nk}^{(0)}$ from LSDA (+$U$) calculations, we take SOC as a
purterbation, and estimate the first-order and second-order\ SOC corrections
wavefunction $\psi _{nk}^{(1)}$ and $\psi _{nk}^{(2)}$ in Wien2k \cite%
{wien2k}. Then all $J_{{}}^{\alpha \beta }$ elements can be calculated
with no need to do the separate self-consistent calculations with different
spin orientations. Meanwhile, this method can produce the first and second order
SOC contribution to the DM and $\Gamma $ interaction separately.

In the following, we will apply our two methods to several typical examples
corresponding to $3d$, $4d$ and $5d$ transition metal oxides respectively in
the next section.

\section{Results}

\subsection{First-principles examples of typical materials}

\begin{table}[tbh]
\caption{The calculated Heisenberg exchange parameters $J$ (in meV)\ for La$%
_{2}$CuO$_{4}$. The calculated spin exchange parameters in the previous
theoretical work are also shown for comparison.}
\label{LaCuO-J}\centering%
\begin{tabular}{cccc}
\hline\hline
& \multicolumn{3}{c}{La$_{2}$CuO$_{4}$} \\ \hline
$J$ & Ref. \cite{wan2009} & Ref. \cite{dm-2005} & this work \\ \hline
$J_{1}$ & 27.2 & 29.2 & 25.76 \\
$J_{2}$ & $-$3.00 & $-$4.1, $-$3.9 & $-$3.80, $-$3.38 \\
$J_{3}$ & $-$0.05 & 0 & $-$0.11 \\ \hline
\end{tabular}%
\end{table}

\begin{table*}[tbh]
\caption{The calculated nearest neighbor DM interaction parameters (in meV)
for La$_{2}$CuO$_{4}$ via the two approaches in this work. $R$ is the radius
vector from two sites of magnetic ions in units of the lattice constant. The
columns $D^{(1)}$ and $D^{(2)}$\ represent the calculated DM interaction
proportional to $\protect\lambda $ and$\ \protect\lambda ^{2}$. Due to the
small SOC, the DM interactions proportional to $\protect\lambda ^{2}$ are
zero with an accuracy of 0.01meV.}
\label{LaCuO-D}\centering%
\begin{tabular}{c|c|cc}
\hline\hline
& \multicolumn{1}{|c|}{WienJ} & \multicolumn{2}{|c}{the second method in this
work} \\ \hline
$R$ & \multicolumn{1}{|c|}{$D$} & $D^{(1)}$ & $D^{(2)}$ \\ \hline
(0.5, $-$0.5, 0) & ($-$0.09, $-$0.14, 0) & ($-$0.09, $-$0.14, 0) & (0, 0, 0)
\\
($-$0.5, $-$0.5, 0) & ($-$0.09, 0.14, 0) & ($-$0.09, 0.14, 0) & (0, 0, 0) \\
($-$0.5, 0.5, 0) & ($-$0.09, $-$0.14, 0) & ($-$0.09, $-$0.14, 0) & (0, 0, 0)
\\
(0.5, 0.5, 0) & ($-$0.09, 0.14, 0) & ($-$0.09, 0.14, 0) & (0, 0, 0) \\ \hline
\end{tabular}%
\end{table*}

\begin{table*}[tbh]
\caption{The calculated DM interaction parameters (in meV) for Ca$_{2}$RuO$%
_{4}$ via the two approaches in this work. Here $R$ is the radius vector
from two sites of magnetic ions in units of the lattice constant.\ The
columns $D^{(1)}$ and $D^{(2)}$\ represent the calculated DM interaction
proportional to $\protect\lambda $ and$\ \protect\lambda ^{2}$. It can be
seen that the calculated interactions via these two approaches are close,
i.e. $D\approx D^{(1)}+D^{(2)}$. }
\label{214D}\centering%
\begin{tabular}{c|c|cc}
\hline\hline
& \multicolumn{1}{|c|}{WienJ} & \multicolumn{2}{|c}{the second method in this
work} \\ \hline
$R$ & $D$ & $D^{(1)}$ & $D^{(2)}$ \\ \hline
(0.5, $-$0.5, 0) & (0.50,-1.03,0.17) & (0.47,-1.19,0.60) & (0.04,0.15,-0.41)
\\
($-$0.5, 0.5, 0) & (0.50,-1.03,0.17) & (0.47,-1.19,0.60) & (0.04,0.15,-0.41)
\\
($-$0.5, $-$0.5, 0) & (-0.43,-0.95,0.17) & (-0.40,-1.10,0.59) &
(-0.04,0.15,-0.41) \\
(0.5, 0.5, 0) & (-0.43,-0.95,0.17) & (-0.40,-1.10,0.59) & (-0.04,0.15,-0.41)
\\ \hline
\end{tabular}%
\end{table*}

\begin{table*}[tbh]
\caption{The calculated DM interaction parameters (in meV) for Ca$_{3}$LiOsO$%
_{6}$ via the two approaches in this work. $R$ is the radius vector from two
sites of magnetic ions in units of the lattice constant. The columns $%
D^{(1)} $ and $D^{(2)}$\ represent the calculated DM interaction
proportional to $\protect\lambda $ and$\ \protect\lambda ^{2}$. }
\label{416D}\centering%
\begin{tabular}{c|c|cc}
\hline\hline
& WienJ & \multicolumn{2}{|c}{the second method in this work} \\ \hline
$R$ & $D$ & $D^{(1)}$ & $D^{(2)}$ \\ \hline
(0.5, 0.5, 0.5) & (0, 0, -0.263) & (0, 0, -0.223) & (0, 0, -0.099) \\
(-0.5, 0.5, 0.5) & (-0.125, 0.205, -0.021) & (-0.083, 0.143, -0.052) &
(-0.055, 0.095, 0.005) \\
(0.5, -0.5, 0.5) & (-0.125, -0.205, -0.021) & (-0.083, -0.143, -0.052) &
(-0.055, -0.095, 0.005) \\
(0.5, 0.5, -0.5) & (0.226, 0, -0.021) & (0.165, 0, -0.052) & (0.110, 0,
0.005) \\ \hline
\end{tabular}%
\end{table*}

A. La$_{2}$CuO$_{4}$

As a benchmark on the accuracy of our methods in calculating Heisenberg and
DM interactions, we first study the famous La$_{2}$CuO$_{4}$, which have
been studied in a number of theoretical work \cite%
{dm-2005,dm-2010,wan2009,lacuo-add-1,lacuo-add-2,yildirim1995anisotropic,thio1988antisymmetric,kastner1988neutron}%
. The LSDA + $U$ ( = 7 eV) \cite{lacuo-U} calculation is applied. Without
SOC considered, the calculated Heisenberg exchange parameters have no
difference between these two approaches, which are summarized in the Table %
\ref{LaCuO-J}. The calculated nearest neighbor magnetic coupling $J_{1}$ are
dominant with the value of about 25.76 meV. We can find that the spin
exchange coupling parameters decrease rapidly with the increasing distance
between two Cu ions.\ The next nearest neighbor magnetic coupling $J_{2}$
shows ferromagnetic behavior and is one order of magnitude smaller than $%
J_{1}$. The third nearest neighbor $J_{3}$ is antiferromagnetic and almost
negligible. The results agree well with the previous theoretical work \cite%
{dm-2005,wan2009}.

The weak ferromagnetism of La$_{2}$CuO$_{4}$ is originated from the canting
of the magnetic moments, which can be descried by the competition of
Heisenberg interaction and DM interaction. Based on the two approaches in
above section, the nearest neighbor\ DM parameters are calculated as shown
in table \ref{LaCuO-D}. As shown in table \ref{LaCuO-D}, the DM interactions
proportional to $\lambda ^{2}$ are negligible due to the small SOC in $3d$
orbital, and the DM interactions proportional to $\lambda $ are almost the
same as the calculated DM parameters in WienJ. According to the calculated
Heisenberg and DM parameters, the value of the canting angle is estimated to
be about 1.7$\times 10^{-3}$, which is in a good agreement with the
experimental value of 2.2-2.9$\times 10^{-3}$ \cite%
{thio1988antisymmetric,kastner1988neutron}. For comparison with previous
theoretical works, Mazurenko et al. \cite{dm-2005} proposed the angle value
of 0.7$\times 10^{-3}$ using Green's function technique. With the
construction of a tight-binding parametrization of the Hamiltonian with SOC,
Katsnelson et al. \cite{dm-2010} calculated canting angle to be 5.0$\times
10^{-3}$. Comparatively our results agree very well with the experiment and
quite promising.

\bigskip

B. Ca$_{2}$RuO$_{4}$

As the example of $4d$ transition metal oxides, Ca$_{2}$RuO$_{4}$
crystallizes in the space group Pbca and has the layered perovskite
structure \cite{Ca2RuO4-1,Ca2RuO4-2,Ca2RuO4-3,Ca2RuO4-4,Ca2RuO4-5,Ca2RuO4-6}%
. The ground state of Ca$_{2}$RuO$_{4}$ is an antiferromagnetic spin
ordering with an insulating electrical behavior \cite{Ca2RuO4-2}. A weak
ferromagnetic component is induced by spin canting below the magnetic
transition temperature 113K \cite{Ca2RuO4-1}.

To study its magnetic properties, we performed the LSDA + $U$ $(=3eV)$ \cite%
{Ca2RuO4-3} calculations for Ca$_{2}$RuO$_{4}$. The calculated nearest
neighbor Heisenberg interaction is about 20.9 meV. Experimentally, the
Heisenberg parameters were estimated via inelastic neutron scattering as 16
meV in Ref. \cite{Ca2RuO4-5} and 5.8 meV in Ref. \cite{Ca2RuO4-6}, and our
result 20.9 meV is closer to the first value. Meanwhile, the calculated DM
interactions by the two approaches mentioned above are both presented in
Table \ref{214D}. As shown in Table \ref{214D}, the calculated DM
interactions in WienJ are also almost the same as the sum of DM interactions
proportional to $\lambda $ and $\lambda ^{2}$, i.e., $D\approx D(\lambda
)+D(\lambda ^{2})$ as shown in Table \ref{214D}. Note that the strength of
first-order SOC corrected DM interactions $\left\vert D(\lambda )\right\vert
$ is around 1.31-1.41 meV, while the $\left\vert D(\lambda ^{2})\right\vert $
is about 0.44 meV, therefore the DM interactions proportional to $\lambda
^{2}$ are non-negligible in such $4d$ magnetic system. The ratio of DM
interactions and Heisenberg interaction is estimated to be $\left\vert
D\right\vert /J\approx 0.05$, which is in good agreement with the rough
estimate 0.06 from experiment \cite{Ca2RuO4-4}.

\bigskip

C. Ca$_{3}$LiOsO$_{6}$

In 5$d$ transition-metal oxides systems, the strength of SOC is expected to
be stronger than 3$d$ or 4$d$ materials due to the large atomic number.
However, in orbital singlet states with relatively large electronic gap such
as\ 5$d^{3}$ with half-filling $t_{2g}$ orbitals \cite{NaOsO3,416-1}, the
electronic structures from fully self-consistent LSDA (+$U$) + SOC
calculation and the ones from further one iteration of SOC calculation after
LSDA (+$U$) calculation have small difference (see Fig. 2 in
Appendix B), indicating that the effect of SOC is still small \cite%
{NaOsO3,416-1}.

As one concrete 5$d^{3}$ example, we focus on Ca$_{3}$LiOsO$_{6}$ \cite%
{416-1,416-2,416-3,416-4} with the crystal structure of K$_{4}$CdCl$_{6}$%
-type. The ground state of Ca$_{3}$LiOsO$_{6}$ is AFM with the magnetic
transition temperature 117K. Both the first-principles study and the
experiment suggested that Ca$_{3}$LiOsO$_{6}$ has a fully opened electronic
gap \cite{416-3}. Though the AFM ordered state has been confirmed
experimentally, the magnetization curve suggests a soft magnetism with a
small spontaneous magnetization. The net magnetization is about 0.02 $\mu
_{B}$ per Os$^{5+}$ ion and is suggested due to a DM interaction generated
by the broken inversion symmetry \cite{416-1}.

We perform the LSDA + $U$ calculations of Ca$_{3}$LiOsO$_{6}$ with $U$ = 2
eV \cite{wan2011topological,NaOsO3,wan2012computational,arita2012ab} and
calculate the magnetic interactions by applying our methods. The Heisenberg
interactions $J_{1}$, $J_{2}$ and $J_{3}$ are estimated to be all AFM with
the values of 13.1 meV, 5.5 meV and 1.1 meV respectively. The $J_{1}$ is the
strongest spin exchange, while $J_{2}$\ is slightly less than one half of $%
J_{1}$, and $J_{3}$\ is an order of magnitude smaller than $J_{1}$. These
properties are in consistent with the energy-mapping results, though our
calculated spin parameters are slightly larger than theirs (9.9 meV, 4.1 meV
and 0.63 meV for $J_{1}$-$J_{3}$\ respectively) \cite{416-4}. Meanwhile, our
numerical\ DM interactions by the two approaches mentioned in Method section
are both summarized in Table \ref{416D}. Since the DM interactions between
the 3rd nearest neighbor and longer-range\ distances for Os$^{5+}$ ions are
negligible, thus we only show the DM interactions for nearest neighbor and
the next nearest neighbor in Table \ref{416D}. It can be seen that the DM
interactions proportional to $\lambda ^{2}$ have the same order of magnitude
as the one proportional to $\lambda $ in Ca$_{3}$LiOsO$_{6}$. According to
the crystal symmetry, the nearest neighbor $D_{1}$ has the form of (0, 0, $%
D_{z}$), and the calculated $D_{1}$ via the two approaches are summarized in
Table \ref{416D}. Meanwhile, there are three different directions of $D_{2}$
connected by the symmetry of threefold rotation along z-axis, as shown in
last three rows of Table \ref{416D}. Summarizing the DM parameters of all
nearest neighbors and the isotropic spin exchange parameters $J$ up to the
3rd nearest neighbor, the expected magnetic moment is estimated to be 0.03 $%
\mu _{B}$, which is in good agreement with the experimental value of 0.02 $%
\mu _{B}$ \cite{416-1}.

\subsection{Materials with the second-order SOC correction to DM interactions%
}

Here we discuss the conditions where the first order of SOC in DM
interactions (i.e. $\mathbf{D}^{(1)})$\ are absent, while DM interactions
proportional to the second order of SOC (i.e. $\mathbf{D}^{(2)})$ are
non-zero. According to the picture of Fig. \ref{toymodeldm} and Eq. (\ref%
{dm1jie}),(\ref{dm2jie}), we summarize these three conditions that need to
be met:\bigskip

\bigskip

(1) the hopping processes between ground state and excited state ($%
t_{mn^{\prime }}$ and$\ t_{m^{\prime }n}$) are symmetry forbidden.

(2) the hopping between ground states ($t_{nn^{\prime }}$) and the hopping
between excited states at two sites ($t_{mm^{\prime }}$) are non-zero.

(3) the relation of orbital angular momentum of the two magnetic ions $%
l_{m^{\prime }n^{\prime }}^{\beta }l_{mn}^{\gamma }-l_{m^{\prime }n^{\prime
}}^{\gamma }l_{mn}^{\beta }$ should be also non-zero.

\bigskip

According to the first condition $t_{mn^{\prime }}=t_{m^{\prime }n}=0$, the
exchange processes in Fig. \ref{toymodeldm}(a) and (c) are forbidden,
therefore $\mathbf{D}^{(1)}$ is constrained to zero. Meanwhile, the second
and third conditions make the exchange processes in Fig. \ref{toymodeldm}(b)
exist, thus $\mathbf{D}^{(2)}$ could be present. Based on these
restrictions, one can easily predict promising candidates with only the
second-order SOC correction to DM interactions according to their different
combinations of crystal symmetry, Wyckoff sites and orbital occupation
pattern. As a simple example, we consider the space group $Pm$ (No. 6) and
put two magnetic ions $A$ and $B$ located at two different $1a$ Wyckoff
Positions $(x,0,z)$ and $(x^{\prime },0,z^{\prime })$. These Wyckoff Positions
have a mirror symmetry. Therefore,
according to the different eigenvalues of mirror operation, we can mark
their states with the irreducible representation A$^{\prime }$ and A$%
^{\prime \prime }$ respectively \cite{bradley2010mathematical}. Here we
assume that their ground states $n$ and $n^{\prime }$ both belong to
representation A$^{\prime }$, while their excited states $m$ and $m^{\prime
} $\ all belong to representation A$^{\prime \prime }$. Thus, the hopping
between the orbitals with different representations are symmetry forbidden,
i.e. $t_{mn^{\prime }}=t_{m^{\prime }n}=0$, while the hopping $t_{nn^{\prime
}}$ and $t_{mm^{\prime }}$\ between the orbitals with same representation
can exist. Moreover, since these two magnetic ions are not related by the
crystal symmetry, $l_{m^{\prime }n^{\prime }}^{\beta }l_{mn}^{\gamma
}-l_{m^{\prime }n^{\prime }}^{\gamma }l_{mn}^{\beta }$ could also be
non-zero. Therefore, this case satisfies all the three conditions we listed
above to be a candidate with only the DM interactions of second order SOC.
The above three\ conditions can be satisfied in many Wyckoff\ positions of
various space groups, and we believe that it may be widespread in a variety
of magnetic materials.

\section{Conclusion}

The magnetic model plays an important role in magnetic investigations. Here
we revisit the general expression of magnetic interactions, including
isotropic exchange interaction, antisymmetric DM interaction and symmetric $%
\Gamma $ term. We clarify that the term proportional to $\lambda ^{2}$ has
both contribution to DM interaction and $\Gamma $ term. We find that the DM
and $\Gamma $ interactions can be separated from their different hopping
processes rather than the orders of SOC. We present two first-principles
methods to calculate the anisotropic magnetic interactions. Based on the
first method, one need perform self-consistent calculations for at least
three different spin orientations to obtain the full nine exchange
parameters $J_{{}}^{\alpha \beta }$. On the other hand, using the second
method, one can estimate these magnetic exchange parameters with no need to
do the separate self-consistent calculations for different spin
orientations. This method can also calculate the first-order and second-order SOC contribution to DM interactions
separately. We have successfully applied our methods to several
typical weak ferromagnetic materials La$_{2}$CuO$_{4}$, Ca$_{2}$RuO$_{4}$
and Ca$_{3}$LiOsO$_{6}$ respectively. Furthermore, according to microscopic
mechanism shown in Fig. \ref{toymodeldm}, we list all three conditions which
can lead to that the DM interactions proportional to $\lambda $ are
symmetric forbidden while the DM interactions proportional to $\lambda ^{2}$
exist.

\section{Acknowledgements}

This work was supported by the NSFC (No. 12188101, 11834006, 12004170,
51721001, 11790311), National Key R\&D Program of China (No.
2022YFA1403601), Natural Science Foundation of Jiangsu Province, China
(Grant No. BK20200326), and the excellent programme in Nanjing University.
Xiangang Wan also acknowledges the support from the Tencent Foundation
through the XPLORER PRIZE.

\bibliographystyle{aps}
\bibliography{DM}

\end{document}